\documentclass[a4paper,fleqn,usenatbib]{mnras}
\usepackage[T1]{fontenc}
\usepackage{ae,aecompl}
\usepackage{latexsym,amssymb}
\usepackage{graphicx}

\title[Jets in Common Envelopes]{Dynamics of jets during the Common Envelope phase}

\author[Moreno M\'endez et al.]{
Enrique Moreno M\'endez$^{1}$\thanks{E-mail: enriquemm@ciencias.unam.mx},
Diego L\'opez-C\'amara$^{2}$\thanks{E-mail: diego@astro.unam.mx},
Fabio De Colle$^{3}$\thanks{E-mail: fabio@nucleares.unam.mx}
\\
$^{1}$Facultad de Ciencias, Universidad Nacional Aut{\'o}noma de M{\'e}xico, A. P. 70-543 04510 D. F. Mexico\\
$^{2}$CONACyT - Instituto de Astronom\'ia, Universidad Nacional Aut{\'o}noma de M{\'e}xico, A. P. 70-264 04510 D. F. Mexico\\
$^{3}$Instituto de Ciencias Nucleares, Universidad Nacional Aut\'onoma de M\'exico, A. P. 70-543 04510 D. F. Mexico}

\date{Accepted XXX. Received YYY; in original form ZZZ}

\pubyear{2017}

\begin{document}
\label{firstpage}
\pagerange{\pageref{firstpage}--\pageref{lastpage}}
\maketitle

\begin{abstract}
Common envelope (CE) is an important phase in the evolution of many binary systems. 
Giant star / compact object interaction in binaries plays an important role in high-energy phenomena as well as in the evolution of their environment.
Material accreted onto the compact object may form a disk and power a jet. 
We study analytically and through numerical simulations the interaction between the jet and the CE.
We determine the conditions under which accreting material quenches the jet or allows it to propagate successfully, in which case even the envelope may be ejected.
Close to the stellar core of the companion the compact object accretes at a larger rate.
A jet  launched from this region needs a larger accretion-to-ejection efficiency to successfully propagate through the CE compared to a jet launched far from the stellar core, and is strongly deflected by the orbital motion.
The energy deposited by the jet may be larger than the binding energy of the envelope. The jet can, thus, play a fundamental role in the CE evolution. 
We find that the energy dissipation of the jet into the CE may stop accretion onto the disk.
We expect the jet to be intermittent, unless the energy deposited is large enough to lead to the unbinding of the outer layers of the CE.
Given that the energy and duration of the jet are similar to those of ultra-long GRBs, we suggest this as a new channel to produce these events.
\end{abstract}


\section{Introduction}
\label{sec:int}
Most massive stars are in binary systems \citep[71\% of O-type stars according to][]{2012Sci...337..444S}.
The interaction between stars in binaries is responsible for producing a variety of  transient phenomena 
in astrophysics, including, among others, type Ia supernovae (single degenerate: \citealt{1973ApJ...186.1007W}; 
double degenerate: \citealt{1984ApJ...277..355W} and \citealt{1984ApJS...54..335I}), gamma-ray bursts 
\citep[e.g.,][]{2007ApJ...671L..41B, 2011ApJ...727...29M, berger14}, ultra-luminous x-ray sources 
\citep[e.g.][]{2005MNRAS.356..401R,2013Natur.503..500L}, novae 
\citep{1956ApJ...123...44C, 2012ApJ...746...61D}, millisecond pulsars \citep{1998Natur.394..344W}, 
gravitational waves \citep{1982ApJ...253..908T, abbott16}, and common envelopes 
\citep[CE;][]{ivanova13}, among others.

The CE phase, in which one of the components of a binary system is engulfed by the stellar envelope of the 
secondary star, is a brief but crucial phase in the formation of many compact binary systems. 
This is especially true in binary systems formed by compact objects with an 
orbital distance smaller than the radius of the progenitors 
\citep[e.g.,][]{1975ApJ...195L..51H, 2003Natur.426..531B}. 
The CE phase generally occurs after a phase of unstable mass transfer where the binary system losses 
angular momentum, or when one of the stars has a fast radius increase due to, e.g., the onset of the 
red-giant phase, and it fills its Roche lobe \citep[RL; see, e.g.,][]{1983ApJ...268..368E}. 

CE phases are short-lived ({\bf $\sim$} a year to a few hundreds of years) 
when compared to other stellar stages.
Thus, observational examples of CEs are scarce.
Nonetheless, \citet{2011A&A...528A.114T} and \citet{2006A&A...451..223T} have suggested that events like those observed in V1309 Sco and V838 Mon, among others, are the result of a CE leading to a merger. Also, \citet{2017ApJ...835..282M} discussed a luminous red nova outburst in M31 as the onset of CE.

Analytical studies of CE evolution are limited because its description involves complicated, unsteady physical processes (e.g., mass transfer and accretion disk physics).
On the other hand, physical phenomena happening at very different scales (e.g., the size of a giant star $\sim  10^{13}$~cm and 
a disk around a compact object $\sim  10^{7}$~cm; \citealt{taam00,ivanova13})
limit the use of numerical simulations.

As a consequence, CE evolution is far from being a well understood problem.
In particular, what brings a CE phase to an end is not well comprehended.
This may be due to the conversion of orbital energy into kinetic energy of the expelled envelope \citep{ivanova13}.
It is also not clear how much mass is accreted by the engulfed object.
This question is especially relevant to determine whether a neutron star (NS) or a black hole (BH)-NS 
binary system are produced after a CE episode between two NS progenitors.

The Bondi-Hoyle-Lyttleton \citep[BHL,][]{bondi44, hoyle39} accretion radius can be a considerable fraction of 
the stellar radius. The accretion rate estimated by \citet{1995ApJ...440..270B} and 
\citet{1996ApJ...459..322C}, when including highly-efficient neutrino cooling, is large enough to favour 
the formation of BH-NS systems over NS-NS systems. However, when one considers the density gradient 
expected inside the envelope of a star, angular momentum prevents accretion by a factor of up to a 
hundred \citep{macleod15,lora15}, likely avoiding the formation of a BH. 

The BHL rate is much larger than the mass accretion rate obtained from the Eddington limit, which considers spherical accretion of material without external pressure. 
However, this limit does not directly apply to accretion through a disk \citep{shakura73}.  
Hypercritical accretion may occur during certain mass transfer stages in binaries with compact objects
\citep{1995PhR...256...95C,1994ApJ...436..843B,2008ApJ...689L...9M,2011MNRAS.413..183M}.  This is especially the case when the mass transfer rate is at least three to four orders of magnitude larger than the Eddington limit for accretion, as it is likely that the energy radiated by the accreted mass will be neutrino dominated.

Numerical simulations are essential to understand the CE. The first CE numerical studies were limited to one- and two- dimensions \citep[e.g.,][]{1984ApJ...280..771B, 1988ApJ...335..862F}. 
Three-dimensional (3D) hydrodynamical studies of the CE phase have also been carried out \citep[e.g.,][]{terman94, rasio96,1988ApJ...329..764L}.
\citet{1998ApJ...500..909S} studied the stages when the envelope is ejected. Also, \citet{2017MNRAS.467.3556B}, found that jet appearance and quenching during CE can produce transient objects. Other 3D studies focused on the formation of degenerate stars \citep{2000ApJ...533..984S}, the formation of Wolf-Rayet stars \citep{2003RMxAC..15...34D}, and the interaction of stellar objects within the CE phase, either by employing Eulerian \citep{2008ApJ...672L..41R, 2012ApJ...746...74R} or Lagrangian \citep{2012ApJ...744...52P} meshes. 

Velocity gradients \citep{ruffert97, ruffert99, armitage00} as well as density gradients \citep{lora15} 
can provide enough angular momentum to form a disk. Studies of jets at different scales (e.g., proto-stellar jets, jets from X-ray binaries, AGN, among others, see e.g., \citealt{livio99}) show that accretion from a disk to a central star is typically associated to the creation of a jet with a velocity of order of the escape velocity and an ejection mass rate of $\sim 10$\% of the accreting mass rate.

The presence of a jet may change the accretion rate onto the disk surrounding the compact object and, hence, the accretion onto the compact object itself. In addition, jets moving through a medium transfer energy to the surrounding environment, and have thus been suggested as an energy-deposition channel which may allow for the removal of the envelope during a CE phase \citep[][]{armitage00,2014arXiv1404.5234S}. 
\citet{2004NewA....9..399S} considered a jet launched by a main-sequence star or a white dwarf inside the envelope of a red giant (RG), while \citet{papish15} considered a jet launched by a neutron star (NS) inside a 16 M$_{\odot}$ RG. Both of them concluded that the jet feedback can eject the entire envelope and even part of the stellar core.

In this paper, we present detailed three-dimensional simulations of the interaction of a jet within the CE. Previous simulations that the jet can be deflected by the relative motion (orbital) of the compact object \citep{2013AN....334..402S}; or, if the jet grazes around the CE, it interacts with the ejected gas and produces hot low-density bubbles \citet{2017MNRAS.465L..54S}.  

The paper is organised as follows. In Section~\ref{sec2} we estimate analytically under which conditions 
a jet can successfully propagate through the CE. Section~\ref{sec:res} presents the 
numerical method and the results of three dimensional adaptive mesh refinement, hydrodynamic simulations.
In Section~\ref{sec:dis} we discuss the results and in Section~\ref{sec:con} we present our conclusions.

\section{Dynamics of the jet propagating through the common envelope}
\label{sec2}

In this Section, we present a simple analytical description of the interaction between the jet and the stellar envelope. 

We consider a compact object moving through the envelope of a RG star. Accretion onto the compact object during the CE phase can lead to the formation of a disk which may drive a jet.
Accretion onto the compact object engulfed inside the CE happens at a 
fraction of the Bondi-Hoyle-Lyttleton (BHL) rate
\begin{eqnarray}
 \dot{M}_{\rm BHL} = \frac{4 \pi G^2 M_{\rm co}^2\rho_\infty}{\left(c_\infty^2+v_\infty^2\right)^{3/2}}\approx
 \frac{4 \pi G^2 M_{\rm co}^2\rho_\infty}{v_\infty^3}
  \;,
\label{eq:bhl}
\end{eqnarray}
where $M_{\rm co}$ is the mass of the compact object, $\rho_\infty$ and $c_\infty$ are the density and 
sound speed of the stellar envelope, $v_\infty$ is the velocity of the compact object with respect to 
the velocity of the stellar envelope. 

We assume that the orbital motion is supersonic, 
i.e. $v_\infty \gg c_\infty$, and place the reference system in co-motion with the compact object. 
Therefore, in this system of reference the compact object is static and a ``wind'' moves 
towards it due to the compact object motion through the envelope of the RG.

The jet dynamics depend on the jet properties at the injection region and on the density stratification 
($\rho_\infty$; see equation \ref{eq:rho}) of the CE. 
If we assume that a fraction of the accreted mass, $\dot{M}_{\rm BHL}$, ends powering the jet, the jet 
kinetic luminosity is directly proportional to the density of the material surrounding the compact object 
(see equation~\ref{eq:bhl}).

\begin{figure}
  \centering
  \includegraphics[width=8 cm]{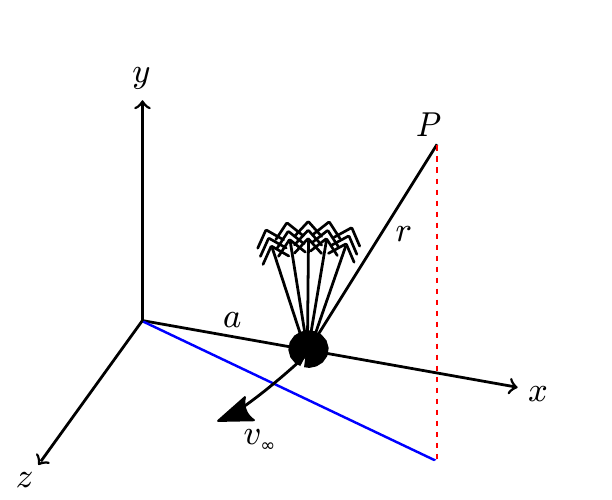}
  \caption{Schematic representation of the reference system used in this paper. $xz$ is the orbital plane. $a$ is the orbital separation between the centre of the donor star and the compact object (located at $x=a$, $y=z=0$). $v_\infty$ is the compact object velocity through the common envelope; $r$ is the distance from the compact object to a point $P$. The jet is assumed to move along the $y$-axis with velocity $v_j$ and opening angle $\theta_j$.}
   \label{fig1}
\end{figure}

Following \citet{papish15}, we consider a 16 $M_\odot$ red giant model \citep{taam78} with a density profile given by
\begin{equation}
 \rho_\infty =  0.68 \times \left(\frac{a}{R_\odot}\right)^{-2.7} \; {\rm g \;cm}^{-3}\;,
\label{eq:rho}
\end{equation}
where $a$ is the distance from the centre of the donor star to the position of the compact object (see Figure~\ref{fig1} for a schematic representation of the geometry of the system).

For this stellar profile, the BHL mass accretion rate (equation~\ref{eq:bhl}) for a $5 M_\odot$ BH is given by:
\begin{eqnarray}
 \dot{M}_{\rm BHL} = 5 \times 10^{29}
 \left( \frac{M_{\rm co}}{5 M_\odot}\right)^2
 \left( \frac{M(a)}{20 M_\odot}\right)^{-3/2}
 \left( \frac{a}{R_\odot}\right)^{-1.2}
  {\rm g\; s}^{-1}
\end{eqnarray}
where $M(a)$ is the mass enclosed within the radius $a$, and we assume that the compact object moves with keplerian velocity $v_\infty \simeq \sqrt{GM(a)/a}$.

Once the jet is ejected from the compact object/disk system, it has to 
make its way through the dense CE material which is falling onto the compact object. 
The jet is ``successful" (i.e. it is able to propagate through the CE) if its ram pressure
is larger than that of the material accreting onto the compact object.
To understand under which conditions this happens, we next estimate the ram pressures of the jet and of the accreting material.

The density of the jet at a distance $r$ from the compact object is given by 
$\rho_j =  \dot{M}_j/\left[ 4 \pi (1-\cos\theta_j)  r^2 v_j \right]$, where $v_j$ and $\theta_j$ 
are the velocity and opening angle  of the jet (measured from the polar axis). Assuming $\theta_j \ll \pi/2$, 
and taking the jet mass ejection rate $\dot{M}_j = \epsilon \dot{M}_{BHL}$ as a fraction $\epsilon$ 
of the BHL mass accretion rate, the jet ram pressure is then given by 
\begin{eqnarray}
 P_j = \rho_j v_j^2= 
\rho_\infty v_\infty^2 
\left(\frac{G M_{\rm co}}{rv_\infty^2}\right)^2
\frac{2 \epsilon v_j}{\theta_j^2 v_\infty}\;.
\label{eq:pj}
\end{eqnarray}

To estimate the ram pressure of the accreting material, we use the approximate description of 
\citet{bisnovatyi79} (see also \citealt{edgar04} and references therein), which studied the 
ballistic (i.e., assuming a ``cold'' fluid) trajectory of the gas falling towards the compact
object. The radial velocity, radius and density of the accretion flow are
\begin{eqnarray}
v_r = - \sqrt{v_\infty^2+\frac{2GM_{\rm co}}{r}-\frac{\zeta^2 v_\infty^2}{r^2}}\;,\nonumber\\
r = \frac{\zeta^2 v_\infty^2}{GM_{\rm co}(1+\cos\theta)+\zeta v_\infty^2 \sin \theta}\;,\\
\rho_a = \frac{\rho_\infty \zeta^2}{r \sin \theta (2\zeta-r \sin \theta)}\;, \nonumber
\end{eqnarray}
where $\zeta$ is the impact parameter and $\theta$ is the polar angle. We can determine the ram pressure of the material falling vertically onto the compact object (i.e., with $\theta=\pi/2$) by inverting these equations. We get
\begin{eqnarray}
P_a = \rho_a v_r^2 = \rho_\infty v_\infty^2
\left(\frac{GM_{\rm co}}{rv_\infty^2}\right)^2
 \frac{1}{\sqrt{1+\frac{4GM_{\rm co}}{rv_\infty^2}}} \;.
\label{eq:pa}
\end{eqnarray}

In the system of reference of the compact object, the CE material streaming sideways acts as a wind 
moving towards the jet and compact object system with  velocity $v_\infty$, 
corresponding to a wind ram pressure:
\begin{eqnarray}
P_w = \rho_\infty v_\infty^2\;.
\label{eq:pw}
\end{eqnarray}

\begin{figure}
  \centering
  \includegraphics[width=8 cm]{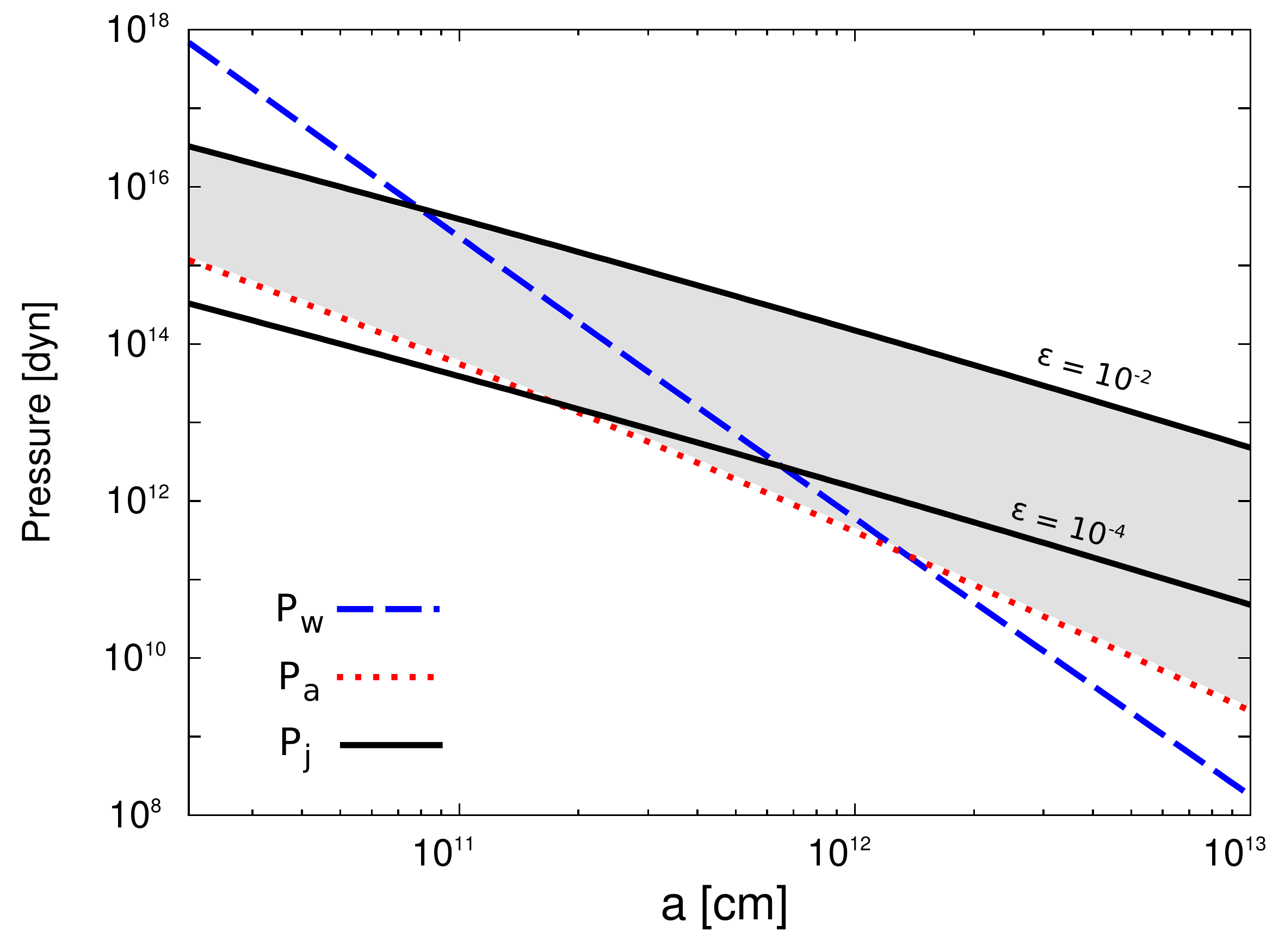}
  \caption{Plot showing the jet propagation scenarios depending on the jet ram pressure $P_{\rm j}$, the 
  ram pressure $P_{\rm a}$ of the material accreting on the compact object/disk system and the ram pressure 
  $P_{\rm w}$ of the CE, which in the system of reference of the compact object is pushing the jet sideways. 
  The shaded region corresponds to the range of parameters where the jet propagates through the 
  common envelope freely ($P_{\rm j}\gg P_{\rm a},P_{\rm w}$) or is deflected 
  ($P_{\rm w}\gtrsim P_{\rm j}\gg P_{\rm a}$). For $P_{\rm j}\lesssim P_{\rm a}$ the jet is quenched. 
  Finally, the jet ram pressure is limited to values of $\epsilon \lesssim 10^{-2}$, where $\epsilon$
  is the fraction of $\dot{M}_{\rm BHL}$ powering the jet.}
   \label{fig2}
\end{figure}

Figure~\ref{fig2} shows the ram pressures (equations~\ref{eq:pj},\ref{eq:pa},\ref{eq:pw}) as a function of the distance $a$
from the centre of the RG star (assuming $\theta_j= 0.1$ radians, $v_j=c/3$, $M_{\rm co}=1$~M$_\odot$, 
and $r=10^{11}$~cm).
The shaded area represents the region of the parameter space where the jet successfully 
overcomes the accreting material ($P_{\rm j} > P_{\rm a}$), within the expected efficiencies 
(i.e., $\epsilon \lesssim 10^{-2}$). Regions above the $\epsilon \approx 10^{-2}$ line are excluded 
as there is not enough mass accreted to the disk to power the jet\footnote{Numerical simulations by 
\citet{macleod15} show that the mass accretion rate $\dot{M}_a$ can be ten to a hundred times smaller 
than $\dot{M}_{\rm BHL}$.}. Observations of astrophysical jets show that typically 
$\dot{M}_j \sim 0.1~\dot{M}_a$. Thus, we assume $\dot{M}_j \lesssim 10^{-2}~ \dot{M}_{\rm BHL}$. 
The kinetic luminosity of the wind determines whether the jet trajectory is substantially 
deflected. The amount of deflection of the jet trajectory also determines the jet energy and 
momentum deposition into the envelope. 

Following the evolutionary track corresponding to $\epsilon = 10^{-2}$ in Figure~\ref{fig2} (with the compact 
object approaching the stellar core with time), the jet initially propagates unperturbed. When the compact 
object moves to radii $a \lesssim 10^{11}$~cm, the jet is substantially deflected by the sideways 
streaming wind. For lower efficiencies (e.g., $\epsilon \lesssim 10^{-4}$), the jet propagates unperturbed 
until $a \gtrsim 10^{12}$~cm. For radial distances $a \lesssim 2 \times 10^{11}$~cm the jet is quenched. 
For very low efficiencies ($\epsilon \lesssim 10^{-5}$) the jet is quenched at all radii.
It is easy to see from equations~\ref{eq:pj} and \ref{eq:pa} that $P_j/P_a$ increases with jet 
collimation (i.e., smaller $\theta_j$), larger accretion/ejection efficiencies $\epsilon$, larger distances $a$ and for CE with lower masses $M(a)$. 
As the ratio $P_j/P_w \propto M_{\rm co}^2$, at a fixed radial distance from the centre of the CE, 
jets from white dwarfs and neutron stars are much more deflected than jets from BH.

In order to better understand the dynamical evolution of the system,  in the next Section we present numerical simulations.

\section{Numerical simulations of a jet inside a CE}
\label{sec:res}

In this Section, we present the numerical setup and numerical simulations of the interaction between a jet launched from a BH, the material accreting onto the BH, and a wind with a density gradient.

In our simulations we assume that the jets are launched perpendicular to the orbital plane (i.e. parallel to the orbital angular momentum). Considering the supernova kick a compact object may receive when formed, this may not always be the case. However, it is a highly likely scenario for a 5 M$_\odot$ BH (the larger mass of the BH gives a lower velocity for similar momenta when compared to a NS, if a SN happens at all). Furthermore, binary interactions after the BH is formed may align the BH spin with the orbit as all mass transfer will occur with the orbital angular momentum. Nonetheless, the dynamics of strongly misaligned jets may be very different from that of our simulations.

\subsection{Simulations setup and input parameters}
\label{sec:input}
We study the dynamics of a jet propagating through a common envelope by performing a series of three-dimensional (3D) simulations. The simulations employ the adaptive mesh refinement, Eulerian code \emph{Mezcal} \citep{decolle12a}, which solves the hydrodynamics equations using a second order shock-capturing solver. The code has been extensively used to study the dynamics of Newtonian 
\citep[e.g.,][]{decolle06, decolle08} and relativistic \citep[e.g.,][]{decolle12b, decolle12c} jets.

For the CE density stratification, we consider a 16 M$_\odot$ red giant star (see equation~\ref{eq:rho}). For the compact object, we consider a BH with a mass M$_{\rm co}=5$~M$_\odot$ located at a radial distance $a=1.1\times10^{13}$~cm from the centre of the RG\footnote{The binary stellar evolution channel leading to this system is similar to those proposed in \citet{2012ApJ...752L...2C} or \citet{2014LRR....17....3P}.}. 

Although our simulations refer to a particular compact object mass, we expect results qualitatively similar for other compact object masses. 
In particular, the results of these simulations can be applied also to the case of a NS engulfed in the envelope of a giant star.

A conical jet with an opening angle $\theta_j$, a velocity $v_j=c/3$ and a density $\rho_j$ is launched from a spherical boundary centred in $(x,y,z) = (0,0,1.1\times 10^{13})$~cm and with a radius r$_{\rm{in}}=10^{11}$~cm (see Figure~\ref{fig1}). 
The jet is injected into the computational box at $t = t_{\rm lag}$ (either at a time large enough for the BHL accretion to achieve a steady state, or at $t=0$). The jet density is initialised as $\rho_j = \eta \rho_\infty v_\infty^2 / v_j^2$, where $\eta =  P_j / P_w$ is the ratio between the jet and wind ram pressures. We vary $\eta$ from 0 to 1000, and we consider models with $\theta_j=15^{\circ}$ and $\theta_j=30^{\circ}$.

The wind material is launched from the $xy$ plane at $z=2\times10^{12}$~cm, with keplerian velocity $v_\infty$.
The simulations are performed in the system 
of reference (SoR) of the compact object. In this SoR, the CE material moves towards the compact object with a velocity $v_\infty$, acting as a wind.
The gravitational effects of the compact object are included in the calculation.
Given that the total accreting mass in the computational box ($\sim 0.01$~M$_\odot$) is much smaller than the mass of the compact object,
we neglect the self-gravity of the CE.

The numerical domain, centred at the compact object position, spans $\pm~2\times10^{12}$~cm in the 
equatorial plane ($xz$) and $2\times10^{12}$~cm along the polar axis ($y$-axis). We take reflection 
boundary conditions across the equatorial plane ($y=0$), inflow boundary condition at the $xy$ plane 
and outflow boundary conditions in the other boundaries. We employ $(64, 32, 64)$ cells at the coarsest level of refinement, with three or 
four levels of refinement, corresponding to a maximum resolution of 
$\Delta x =\Delta y=\Delta z =7.8/15.6\times 10^{9}$~cm for the high- and low- resolution models 
(labelled with ``HR'' and ``LR'' in Table~\ref{table1}) respectively. The base of the jet and the 
surrounding region ($r \leq 3 \times10^{11}$~cm) is resolved at the maximum resolution, while the 
jet cocoon by lower levels of refinement. 
The total integration time is either {0.2, 1.4, or 
10}$\times 10^{5}$~s (the wind crossing time is 10$^5$~s) depending on the model. The values of $\eta$, t$_{\rm{lag}}$, jet opening 
angle $\theta_j$, resolution and integration time t$_{\rm int}$ for all the models are shown in Table~\ref{table1}.

\begin{table}
\caption{Initial conditions of the numerical model}
\begin{center}
\begin{tabular}{cccccc}
  \hline
  Model & $\eta=P_j/P_w$ & t$_{\rm{lag}}$(s) & $\theta_j$ & Resolution & t$_{\rm{int}}$(10$^5$~s)\\
  \hline
  BHLhr & 0 & 0  & 30 & HR & 10.0 \\
  BHLlr & 0 & 0  & 30 & LR & 10.0 \\     
  small hr & 0 & 0  & 30 & HR & 10.0 \\ 
  small lr & 0 & 0  & 30 & LR & 10.0 \\ 
  $\eta_{100}$ & 100 & 10$^5$  & 30 & HR & 1.4 \\
  $\eta_{200}$ & 200 & 10$^5$  & 30  & HR & 1.4 \\
  $\eta_{225}$ & 225 & 10$^5$  & 30 & HR & 1.4 \\
  $\eta_{250}$ & 250 & 10$^5$  & 30 & HR & 1.4 \\
  $\eta_{300}$ & 300 & 10$^5$  & 30 & HR & 1.4 \\
  $\eta_{500}$ & 500 & 10$^5$  & 30 & HR & 1.4 \\         
  $\eta_{1000}$ & 1000 & 10$^5$  & 30 & HR & 1.4 \\ 
  narrow & 300 & 10$^5$  & 15 & HR & 1.4 \\
  nolag & 100 & 0  & 30 & HR & 0.2 \\        
  \hline
\end{tabular}
\end{center}
\label{table1}
\end{table}

\subsection{{\bf Accretion on the compact object}}
\label{sec:bhl}
To test the implementation of our initial conditions, we simulate the accretion onto a compact object in the case when there is not a jet injected in the computational domain, and including a density gradient in the accreting material.

\begin{figure}
  \centering
  \includegraphics[width=9 cm]{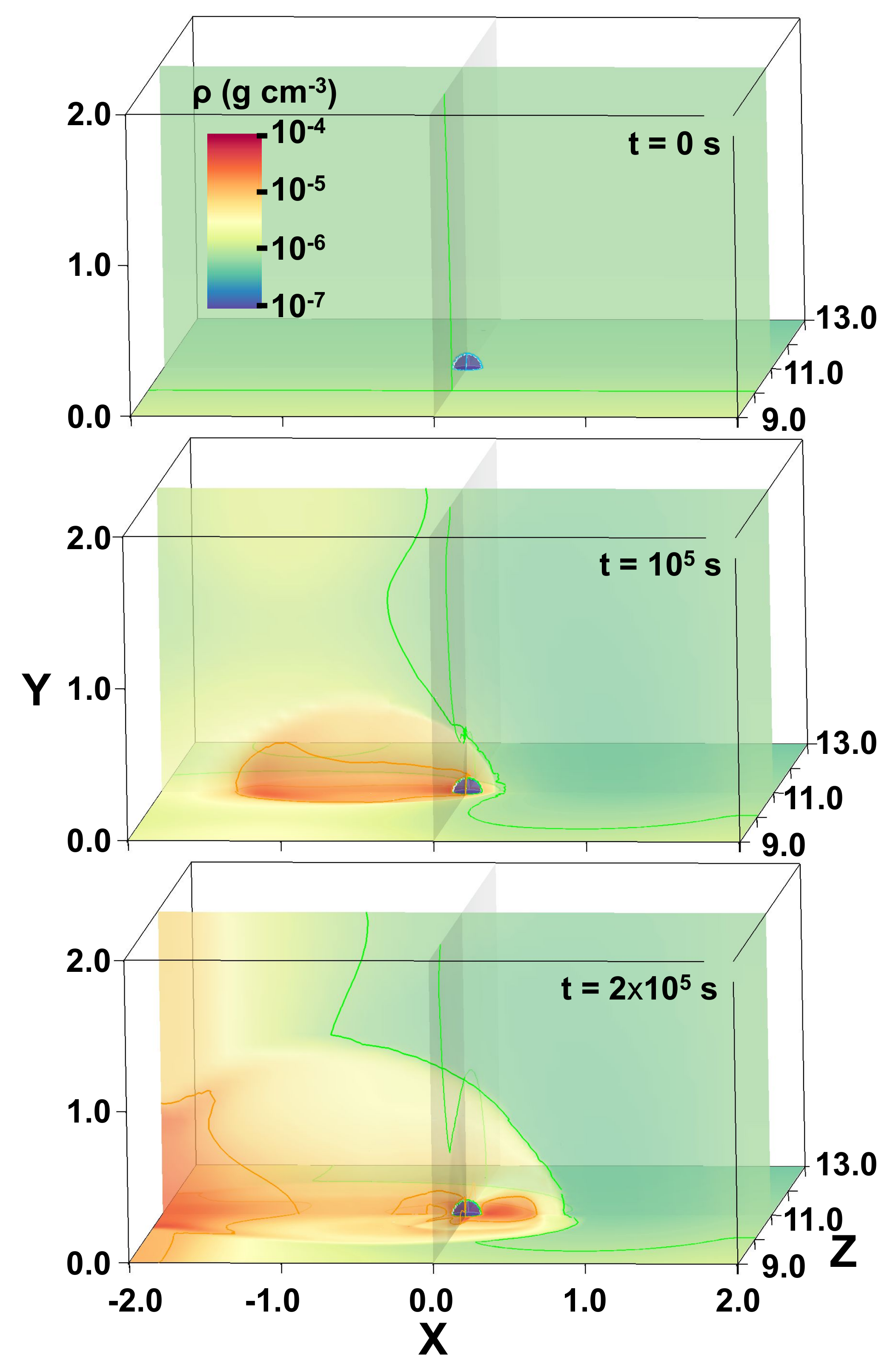}
  \caption{$xz$, $xy$, and $yz$ density map slices showing the time evolution of the
   accretion onto the compact object ($\eta=0$, model BHLhr, see Table~\ref{table1}). The axis are in units of $10^{12}$~cm. 
  The orange and green isocontour lines correspond to a density $10^{-5}$, $10^{-6}$~g~cm$^{-3}$, respectively. An animation (fig3BHLhr.mov) of this figure is available in the online journal.}
  \label{fig3}
\end{figure}

Figure~\ref{fig3} shows the temporal evolution of the accretion on the compact object (model BHLhr).
The shocked material accreting onto the compact object forms an elongated structure (hereafter, 
the ``bulge'') which after approximately 10$^{5}$~s completely engulfs the BH and has density values of 
order $\rho \sim 10^{-5}$ g cm$^{-3}$, this is, $\sim 10$ times larger than the density of the ambient medium 
at that radius. Due to the drop in the CE density as a function of $z$ (see
 equation~\ref{eq:rho}), accretion onto the compact object is asymmetric. The excess in the angular momentum 
 results in the formation of a disk-like structure at $t=2\times 10^{5}$~s, with 
 a radius slightly smaller than 5$\times$10$^{11}$~cm, and 
density values close to 10$^{-6}$ g cm$^{-3}$. The latter can be clearly seen in the lower panel of 
Figure~\ref{fig3}.

To check for numerical artefacts, we also ran a simulation with the same input and boundary conditions as 
model BHLhr, but with lower resolution (three levels of refinement, BHLlr), and two simulations 
with a smaller computational domain (reducing the size of each axis by a half), for both high and low 
resolutions (models small hr and small lr, respectively). 

\begin{figure}
  \centering
  \includegraphics[width=8 cm]{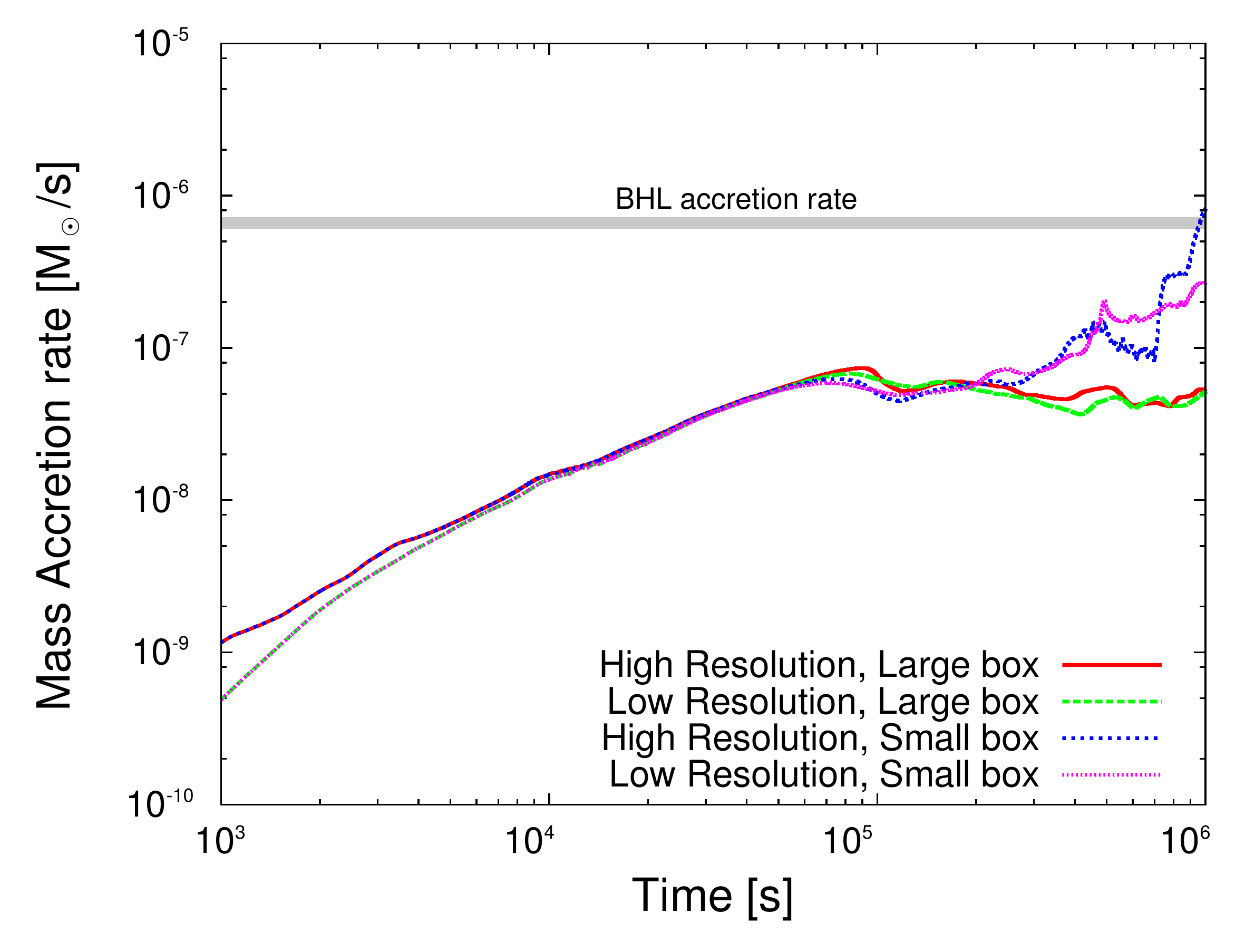}
  \caption{
Mass accretion rate onto the compact object for different resolutions and different domain sizes. The grey solid line corresponds to the analytic solution of BHL.}
  \label{fig4}
\end{figure}

The temporal evolution of the mass accretion rate $\dot{M}$ at the inner boundary $r_{\rm{in}}$ is 
shown in Figure~\ref{fig4}. Independently of the resolution or domain size $\dot{M}$ increases until 
$t\sim 10^5$~s, when the accretion rate achieves a quasi-steady state with 
$\dot{M} \sim 10^{-7}$ M$_\odot$ s$^{-1}$ $\sim 1$ M$_\odot$ yr$^{-1}$ at $t \sim 10^5$~s. 
This accretion rate is much smaller than $\dot{M}_{\rm BHL}$ 
($\dot{M} \approx 0.1\; \dot{M}_{\rm BHL}$, see the grey line in Figure~\ref{fig4}) because of the presence of a 
density gradient in the CE density profile, and thus in the accreting material. 

These results are qualitatively consistent with those of \citet{macleod15}, 
and independent of the resolution when using a large domain.
In particular, the mass accretion rate estimated from this the simulation
(see Figure~\ref{fig4}) is slightly larger ($\approx 50\%$) with respect to the value
determined by \citet{macleod15} (see their Figure 10).

Figure~\ref{fig4} shows that the size of the computational box must be large enough to avoid boundary 
effects which strongly affect the BHL accretion at late times.
In the rest of the simulations we employ four levels of refinement and the large domain size 
(as in the simulation shown in Figure~\ref{fig3}).

\subsection{Global morphology of a jet propagating in a CE}
\label{sec:global}
In order to understand the dynamics of a jet launched from the disk around a compact object within a CE, 
we ran a series of numerical simulations with different values of $\eta = P_j/P_w$ (see Table \ref{table1}). 
As discussed in Section~\ref{sec:bhl}, a quasi-steady state in the accretion process on the compact object 
is achieved after $\sim 10^5$~s, when the accreted material has formed a bulge that covers the BH. 
Thus, to properly model the interaction between the jet and the accreting material, the jet is injected 
after a time $t=t_{\rm lag}=10^5$~s.

\begin{figure}
  \centering
  \includegraphics[width=9 cm]{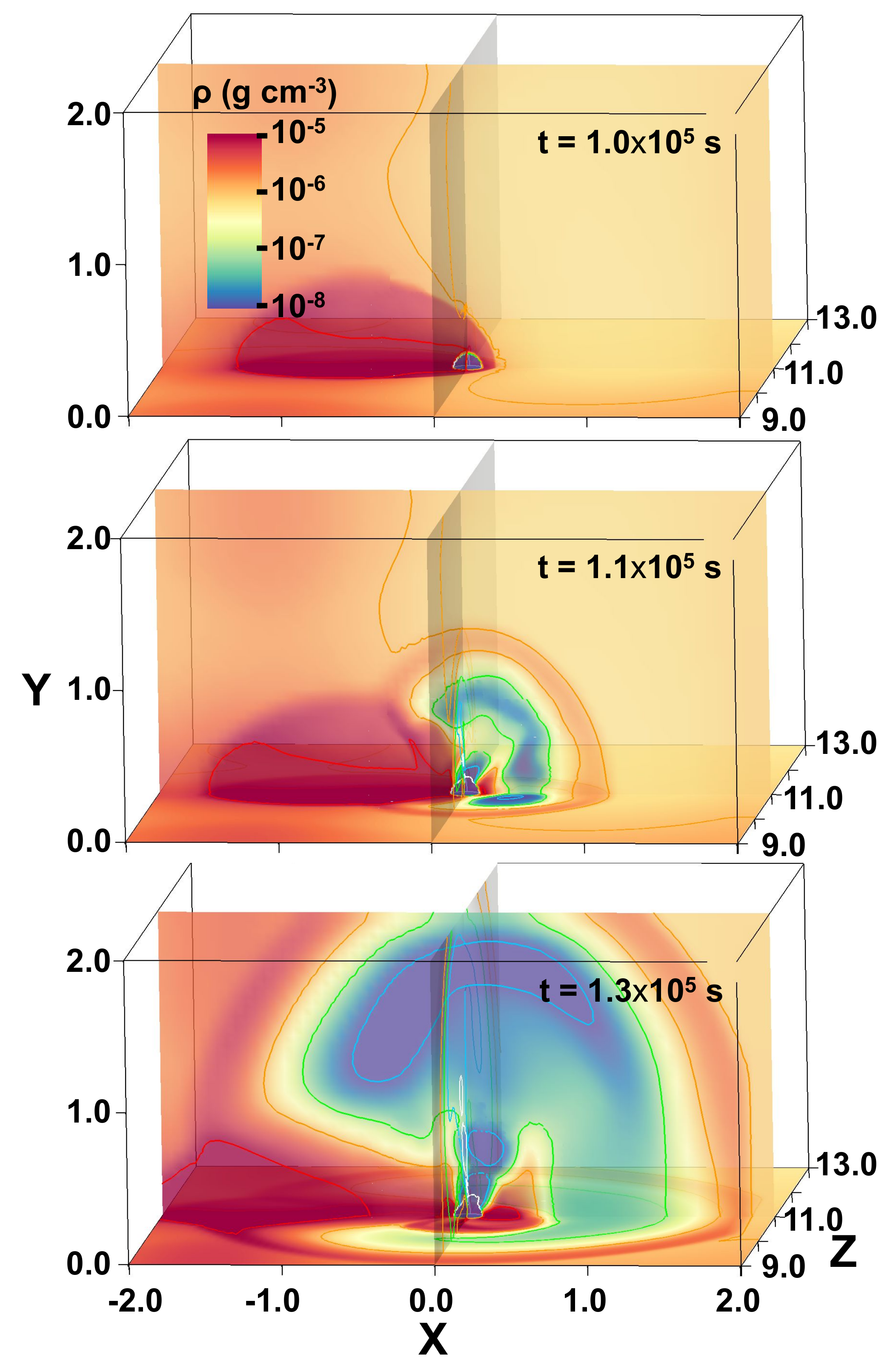}
  \caption{$xz$, $xy$, and $yz$ density map slices showing the evolution of a delayed jet with $\eta=300$ 
  (model $\eta_{300}$, see Table \ref{table1}). The axis are in units of $10^{12}$~cm. 
  The red, orange, green, cyan, and white isocontour lines 
  correspond to $10^{-5}$, $10^{-6}$, $10^{-7}$, $10^{-8}$, $10^{-9}$ g cm$^{-3}$, respectively. An animation (fig5eta300.mov) of this figure is available in the online journal.}
   \label{fig5}
\end{figure}

Figure~\ref{fig5} shows the temporal evolution of a jet with $\eta=300$. The jet first drills through the bulge (i.e., the shocked material accreting onto the compact object) before moving through the 
stellar envelope with a velocity considerably 
slower than the injected velocity ($v\sim 0.01 \;v_j$) 
and reaches $y\sim10^{12}$~cm at $t\sim1.1\times10^5$~s.
Interestingly, the bulge pressure and density gradients push the jet against the wind 
(i.e. in the compact object orbital direction). 
As the jet breaks out of the accreting structure (central, bottom panels of Figure~\ref{fig5}), it expands 
and accelerates into the CE environment. The stellar material (accelerated and heated by the forward shock), 
and the jet plasma (decelerated by the reverse shock), expand sideways forming a cocoon 
which by $t \sim1.3\times10^5$~s has covered the bulge completely.

\subsection{Varying the jet ram pressure}
\label{sec:etas}
To understand how the pressure of the jet (compared to that from the accreted material) 
affects the morphology of the jet, we performed numerical simulations of jets injected with different 
$\eta$ values.

\begin{figure}
  \centering
  \includegraphics[width=9 cm]{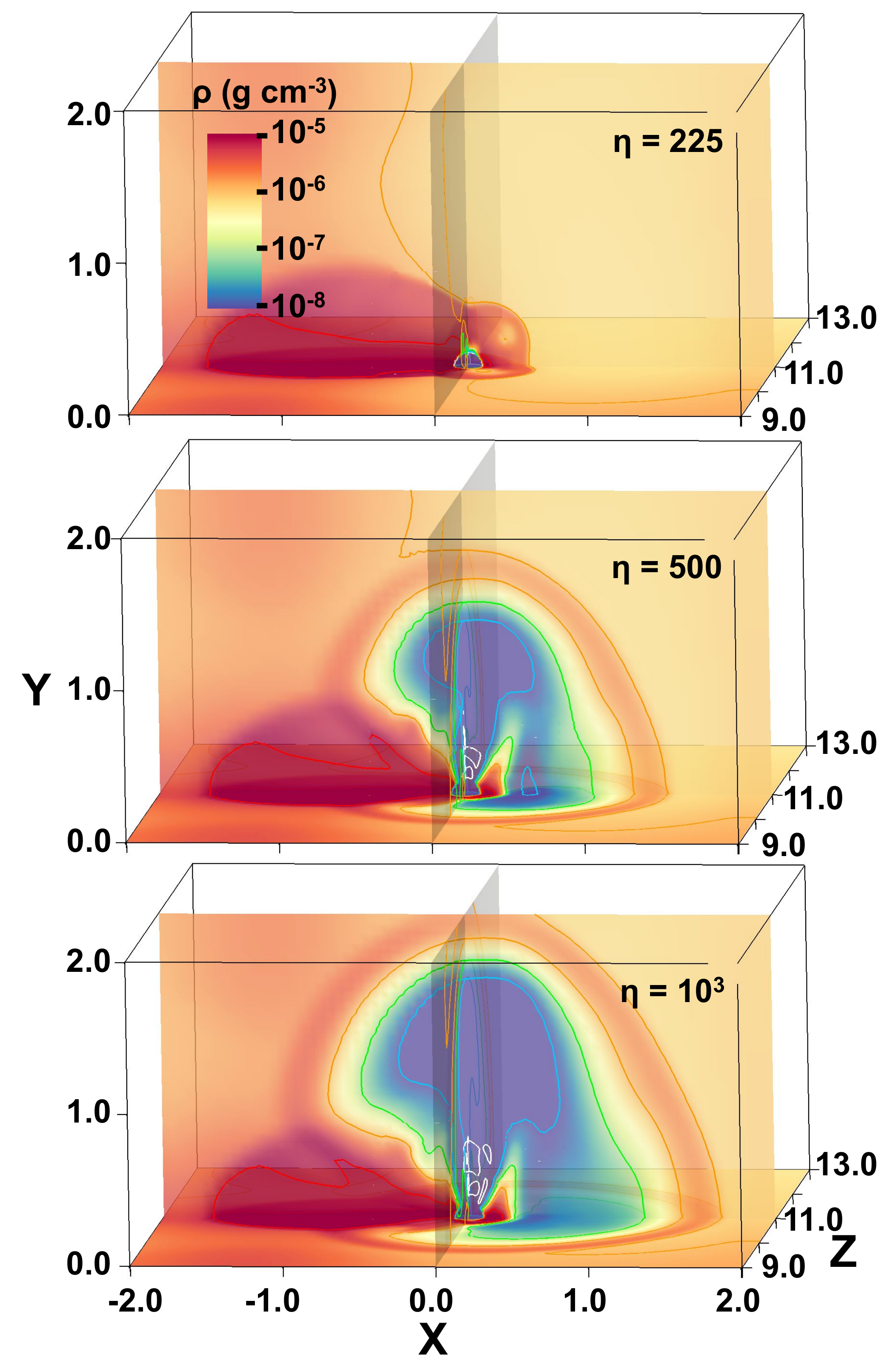}
  \caption{Same as Figure \ref{fig5} for delayed jets with different values of $\eta=P_j/P_w$ (all shown at $t=1.1\times 10^5$~s). Upper panel: $\eta=225$, middle panel: $\eta=500$, and bottom panel: $\eta=1000$ (models $\eta_{225}$, $\eta_{500}$ and $\eta_{1000}$ of Table \ref{table1}, respectively).
   Animations (fig6eta225.mov, fig6eta500.mov, fig6eta1000.mov) of this figure are available in the online journal.}
   \label{fig6}
  \end{figure}

Figure~\ref{fig6} shows the morphology of three jets with different ram pressures 
($\eta=P_j/P_w=225$, 500, and 1000) all at the same time ($t=1.1\times 10^5$~s) 
and with the rest of the jet parameters and boundaries identical to those of the $\eta=300$ model 
(see Section~\ref{sec:global} and Table \ref{table1} for more details). 
As discussed in Section~\ref{sec2}, the jet needs a ram pressure larger than that of the material 
accreting onto the compact object in order to successfully propagate through the stellar envelope. 

From our set of simulations, we find that for the chosen input parameters a jet which is injected with a ram pressure 
$P_j \lesssim 225  \; P_w$ is quenched (as, at this radius, $P_a \gg P_w$, see figure~\ref{fig2}). In this case, the evolution of the system is similar to the case of BHL accretion discussed in Section~\ref{sec:bhl}. 
On the other hand, if the ram pressure of the jet is $P_j \gtrsim 225 \; P_w$, then the jet is able 
to propagate through the bulge and stellar envelope. As expected, simulations with larger jet kinetic luminosities 
(corresponding to larger $\eta$ values) move through the bulge and stellar envelope faster and 
deposit a larger amount of energy into the cocoon. 
Nevertheless, the global morphology of the jet and cocoon is basically independent of $\eta$ 
as long as it is able to drill through the bulge. 

\subsection{Jet opening angle and jet launched into an unperturbed medium}
\label{sec:angles}
Lastly, we study how the jet collimation and launching time affect the morphology of the jet by performing a set of numerical simulations with a narrower 
jet ($\theta_j = 15^\circ$, model ``narrow'') and a jet launched without waiting for a steady state in the 
accretion rate to be achieved ($t_{\rm lag} =0$~s, model ``nolag'').

\begin{figure}
  \centering
  \includegraphics[width=9 cm]{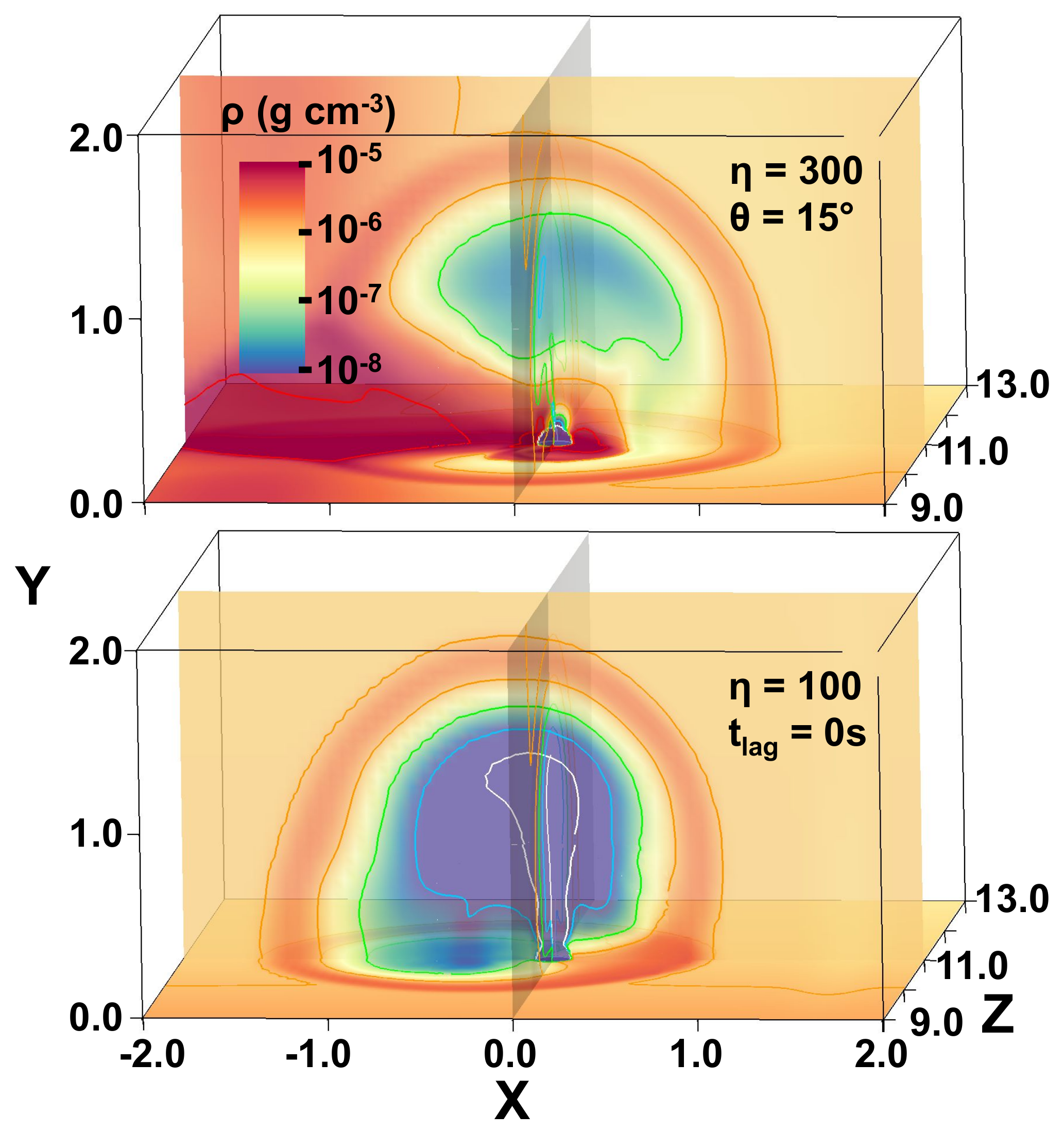}
  \caption{Same as Figure \ref{fig5} for the models $\theta_{15^\circ}$ (bottom panel) and ``nolag'' (bottom panel) of Table~\ref{table1}. The upper panel corresponds to a delayed jet with an opening angle $\theta_j=15^{\circ}$ at $t=1.3\times 10^5$~s. The bottom panel shows a jet (at the evolutionary time $t=1.8\times 10^4$~s) injected at $t_{\rm lag}=0$, i.e. without waiting for the formation of a steady-state accretion structure around the compact object.
  Animations (fig7narrow.mov, fig7nolag.mov) of this figure are available in the online journal.}
  \label{fig7}
\end{figure}

The upper panel of Figure~\ref{fig7} illustrates the case of the narrow jet model. As the injected energy 
is smaller, the jet needs a larger value of $\eta$ to be able to successfully drill through the bulge and 
moves slower through the stellar medium. 
In this case, as long as the jet has $\eta\gtrsim 250$, the 
evolution and the global morphology (the jet and cocoon densities, and the structure of the bulge) 
is alike the successful jet models with a larger opening angle presented in Section \ref{sec:etas}. 

The bottom panel of Figure~\ref{fig7} shows the evolution of the nolag model in which the jet is launched at 
$t_{\rm lag}=0$. 
This scenario corresponds to the case in which the compact object has an accretion disk and a jet when the CE engulfs the compact object 
\citep[e.g.,][]{2016NewAR..75....1S, 2017MNRAS.465L..54S}. 
A much lower kinetic luminosity is needed for the jet to move successfully into the CE, 
as it does not have to push sideways the material accreting onto the compact object (i.e. the bulge 
observed in  the other simulations). 

The nolag model in Figure~\ref{fig7} shows the morphology of a jet with $\eta=100$ 
at $t=1.8\times 10^4$~s. Comparing such morphology with any other model from Figure~\ref{fig5} or 
Figure~\ref{fig6} it is clear that the lack of the bulge drastically modifies the evolution of the jet/cocoon. 
The jet is less dense ($\sim$10$^{-9}$~g~cm$^{-3}$) and propagates $\sim$10\% faster. The cocoon is also 
less dense ($\sim$10$^{-8}$~g~cm$^{-3}$), and quenches the mass accretion rate onto the compact object. 
Since the jet breaks out of the stellar envelope in a shorter time, it deposits less energy in 
the envelope with respect to the other jet models. 
Although the jet ram pressure is much larger than the wind ram pressure, the interaction between the 
jet and the wind slightly bends the jet in the opposite direction with respect to the compact object orbit.

\section{Discussion}
\label{sec:dis}

Our analytical results (Section~\ref{sec2}) are qualitatively consistent with the results of our numerical simulations (Section~\ref{sec:res}).
If the kinetic luminosity of the jet is larger than the ram pressure of the material accreted on the compact 
object, the jet propagates successfully through the CE.

\begin{figure}
  \centering
  \includegraphics[width=8 cm]{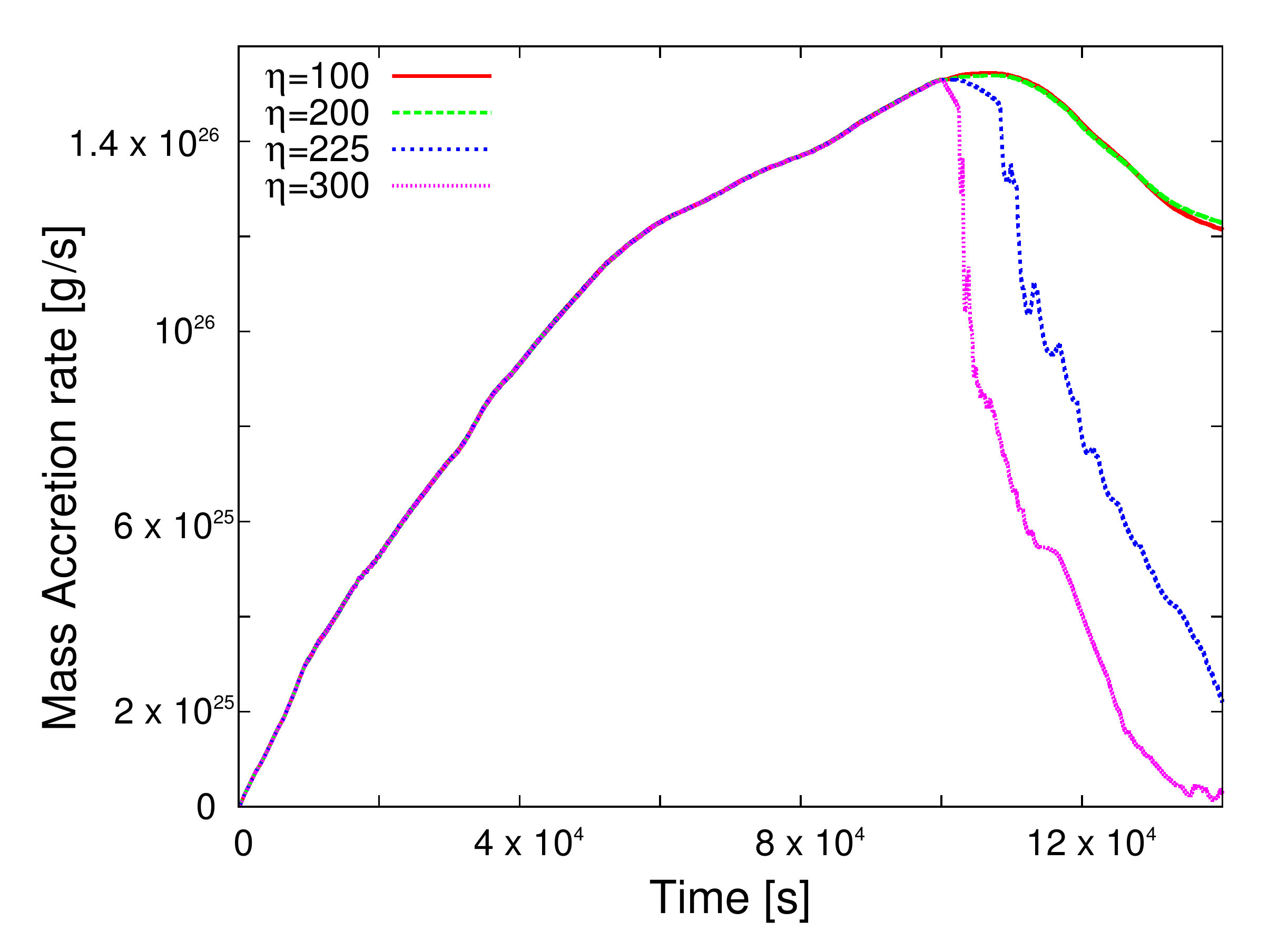}
  \caption{
Mass accretion rate onto the compact object for different values of $\eta=P_j/P_w$ (see Table \ref{table1}). 
Increasing the jet luminosity (i.e., for larger values of $\eta$) the energy deposited in the cocoon increases
and the mass accretion rate on the compact object drops. For values of $\eta \lesssim 225$ the jet is drowned
by the accreting material.}
  \label{fig8}
\end{figure}

Nevertheless, we find that the jet dynamics is considerably more complex than what can be predicted analytically. 
The jet has to move, first, through the asymmetrical bulge (the material accreting onto the compact object). 
Due to the large pressure and density gradients in the region behind the BH, the jet is tilted forwards 
in the direction of motion of the compact object within the CE (the ``wind'' or upstream).
During its propagation, the jet deposits energy into a cocoon, which expands mainly upstream with a velocity of order of a half of the jet velocity.
Downstream the cocoon expands at a much lower speed as it collides 
with the dense bulge structure accreting on the compact object. 
The most important effect of this interaction is the abrupt decline of the mass accretion rate (by an order of magnitude in $\sim 2\times 10^4$~s) onto the compact object (see Figure~\ref{fig8}) due to the ram pressure of the cocoon 
\citep[what is referred to as negative jet feedback mechanism by][]{2016NewAR..75....1S}.
The drop in the accretion rate due to the expanding cocoon implies that the formation of black holes from neutron stars is unlikely, in the
presence of a jet.

The jet is powered by a fraction of the accreted material which is ejected from the inner region of the 
disk/compact object system. 
As the mass accretion rate drops due to the cocoon expansion, the jet is shut down within a viscous 
timescale, $t_{\rm visc}= 1.2/(\alpha \Omega) \left(R/H\right)^2$, \citep{shakura73}:
\begin{equation}
  t_{\rm visc}    \approx 10^6 \;\left( \frac{\alpha}{0.1}\right)^{-1} \left(\frac{R}{10^{10} \;{\rm cm}}\right)^{3/2}\left(\frac{M_d}{0.01\; M_\odot} \right)^{-1/2} {\rm s}\;,
\end{equation}
where $R$ and $H$ are the accretion disk radius and vertical size (assumed to be $H=0.1 \;R$), $M_d$ is the mass of the disk, $\alpha$
is the accretion efficiency and $\Omega$ is the keplerian disk angular velocity. Thus, we can expect that the jet has a lifetime of a few days to weeks.

It has been suggested that the energy deposited by the jet can be large enough to unbind the 
envelope and terminate the CE phase \citep{armitage00, 2004NewA....9..399S, 2014arXiv1404.5234S, papish15}. 
The binding energy of the stellar core and the envelope are 
$E_{\rm b,core} \approx GM_c^2/R_c =  2\times10^{50}$~ergs and $E_{\rm b,layer} \sim4\times 10^{49}$~ergs 
\citep{papish15} respectively.
Assuming that the radius of the CE is $R_\star = 10^{13}$~cm and the jet is launched with a velocity 
$v_{\rm jet} = c/3$, the break-out time can be estimated as  \citep{bromberg11}:\\
\begin{eqnarray}
t_{\rm bo} \simeq 5\times10^4\left(\frac{v_{\rm jet}}{c/3}\right)^{-2/3}\left(\frac{\theta}{30^\circ}\right)^{2/3} \left(\frac{M_\star}{16M_\odot}\right)^{5/6}\nonumber \\ 
\left(\frac{\epsilon}{10^{-3}}\right)^{-1/3}\left(\frac{R_\star^2-a^2}{10^{26}{\rm cm^2}}\right)^{1/3}\left(\frac{a}{10^{13}{\rm cm}}\right)^{2/5} \left(\frac{M_{\rm co}}{5M_\odot}\right)^{-2/3}\; {\rm s} \;,
\end{eqnarray}
which is consistent with the simulation (see, e.g., bottom panel of Figure~\ref{fig5}). 
Given the jet luminosity  ($L_j = \dot{M}_j v_j^2$)
\begin{eqnarray}
 L_j \simeq 10^{44}
  \frac{\epsilon}{10^{-3}}
 \left( \frac{M(a)}{20 M_\odot}\right)^{-3/2}
 \left( \frac{M_{\rm co}}{5 M_\odot}\right)^2 \nonumber\\
 \left( \frac{a}{10^{13}\; \rm cm}\right)^{-1.2}
 \left( \frac{v_j}{c/3}\right)^{2}
  {\rm erg\; s}^{-1} \;,
\end{eqnarray}
the energy deposited by the jet in the cocoon is $E_{\rm jet} = L_j t_{\rm bo}$
\begin{eqnarray}
E_{\rm jet} \simeq 5  \times 10^{48}
\left(\frac{a}{10^{13}{\rm cm}}\right)^{-\frac{4}{5}}\left(\frac{R_\star^2-a^2}{10^{26}{\rm cm^2}}\right)^{1/3}  \left(\frac{\theta}{30^\circ}\right)^{2/3}
\left(\frac{v_{\rm jet}}{c/3}\right)^{4/3}
\nonumber \\
 \left(\frac{\epsilon}{10^{-3}}\right)^{2/3} \left(\frac{M_{\rm  co}}{5M_\odot}\right)^{4/3}
 \left(\frac{M_\star}{16M_\odot}\right)^{5/6} \left(\frac{M(a)}{20M_\odot}\right)^{-3/2} {\rm erg} \;,
\label{eq20}
\end{eqnarray}
which is comparable to that from \citet{2016A&A...596A..58K}.

Successful jets produced at large stellar radii can have a large range of values of accretion to ejection 
efficiencies ($\epsilon = 10^{-5}-10^{-2}$; see Figure~\ref{fig2}). 
If the efficiency is low, the energy deposited by the jet in the envelope is smaller than the 
binding energy. 
Thus, in this case the CE envelope is not unbound by the jet. 
The jet is then switched off by the lack of accreting material. If the cocoon expands and dissipate its energy into the CE, 
then the jet likely re-appears once material surrounding the BH falls back and creates a new accretion disk in a dynamical timescale $t_{\rm ff} \sim 10^5-10^6$~s, from which the jet is once more powered. 
The jet is intermittent in this case, alternating periods of accretion/ejection to/from the compact 
object to quiescent periods of expansion/collapse of the material surrounding the compact object.

A very different outcome is expected if the jet is ejected at large orbital separation and with large 
efficiency ($\epsilon \gtrsim 10^{-4}-10^{-3}$) or at small orbital separation (large values of $\epsilon$ 
are expected in this case for successful jets, see Figure~\ref{fig2}).
In these cases, the cocoon energy (see equation~\ref{eq20}) is large enough to unbind a substantial 
fraction of the CE material.
Furthermore, at smaller orbital separation the jet is strongly deflected
by the larger ram pressure of the wind. 
As the orbital period of the binary system is much smaller than the lifetime and crossing time of the jet (i.e. $P = 4 \times10^3 a_{11}^{3/2} M_{21} $ s $ \ll t_{\rm visc}$), 
the jet deposites a larger amount of energy ($\sim t_{\rm visc}  E_{\rm jet}/t_{\rm bo} \gg E_{\rm jet}$).

Due to numerical limitations, in our simulations we considered relatively large jet opening angles. 
In the case of a more collimated jet, the cocoon may be tilted enough to allow accretion on a narrow 
equatorial region where material may still feed a disk thus allowing for accretion disk, jets and their 
respective cocoons to coexist in a stable phase. 

Althought in this paper we have considered only the case of a jet created during the CE, a different scenario would be that of a pre-existing disk/jet formed during the mass -transfer phase. Such a disk would be much larger in size and less dense. The jet will only interact with the outskirts of the CE in this case. On the other hand, it is unclear whether the disk would survive the onset of a CE for long time.

A jet ejected at $a\lesssim 10^{11}$~cm with a relatively large efficiency ($\epsilon \gtrsim 10^{-2}$) 
deposits an amount of energy large enough to unbind the outer layers of the star. The
jet then propagates freely in the interstellar medium. For such large efficiencies 
and radial distances the jet may reach luminosities of order $10^{51} - 10^{51}$~erg~s$^{-1}$. 
Thus, taking into account that the jets timescale is $\lesssim$ than the viscous timescale, we speculate that this can be an alternative channel to produce highly relativistic jet as those observed associated to ultra-long GRBs \citep{2014ApJ...781...13L}.

When the jet and the cocoon break out of the CE, an X-ray flash would be expected. The energy deposited 
by the relativistic jet into the fast-moving expanding envelope (the cocoon) should produce a quasi-thermal
 emission peaking in optical and UV bands, associated to the emission generated from the relativistic jet 
 which may or may not point to the observers line of sight. If it does, then a hard spectrum would be 
 expected. Otherwise, an off-axis jet could produce a radio signal from the interaction with the environment.
Furthermore, if the jet is launched close to the stellar core of the companion and enough energy is deposited close to it, the core could be disrupted and a SN-like (type II) signal could be produced.

\section{Conclusion}
\label{sec:con}
In this paper we presented three dimensional, hydrodynamic simulations of the interaction between a jet launched from a compact object and the CE. We showed that jets can play a fundamental role in the CE phase.

We find that, in absence of a jet, the accretion rate onto the compact object is a fraction of the BHL rate. 
Also, we have established the conditions for a jet to be able to traverse through the bulge and the stellar CE:
the accretion-to-ejection efficiency (i.e., the rate of accreting material which is converted into an outgoing jet). Depending on the position of the compact object within the engulfing star, this efficiency has to be larger than $10^{-5}$ to $10^{-3}$ and the jet luminosity must be larger than $\sim10^{45}$~erg~s$^{-1}$.

The jet deposits a substantial amount of energy ($E \sim 10^{50}$~erg) into a hot, dense cocoon, which 
inhibits mass accretion on the compact object, and it is likely to affect substantially the CE evolution.
The timescale for disrupting the outer layers of the CE may be substantially shorter than the viscous timescale for the material 
in the accretion disk, thus, even if accretion is suppressed, the jet is not immediately quenched.
This behaviour may lead to three possible scenarios. 1) The jet may be intermittent, as it may push 
away the accreting material, but eventually restart as the material fall back on the compact object. 2) 
It may possibly reach a steady state if the jet is very collimated.  3) It may deposit so much energy 
into the CE that it unbinds its outer layers (if the accretion to ejection efficiency is $\gtrsim 10^{-3}$). 
However, the donor star may eventually fill its Roche lobe again and lead to another CE phase.  

It has been suggested that, during a CE phase, a binary can substantially decrease its period (and separation) 
by converting orbital kinetic energy into kinetic energy of the outer layers, i.e., unbinding the envelope. 
As the energy deposited by the jet in the envelope is comparable to the orbital energy dissipation 
\citep{1976IAUS...73...35V}, it should be taken into account in the CE energy-balance and may even 
terminate the CE phase. Also, the characteristic luminosity, duration and environment of a jet ejected 
from a compact object orbiting at small distances from the center of the CE imply that this can be a 
possible channel to explain some of the ultra-long GRBs.

Local simulations (e.g., which study only a portion of the CE) as those presented here do not includes 
the full CE phenomenology, e.g. the disk formation by accretion on the compact object or the survival of 
a pre-existent accretion disk when the CE phase begins, and the long-term dynamical interaction of the 
jet with the CE and companion star. 
Nevertheless, we believe that this study represents a step in understanding the key role that jets, if 
present, may play in the evolution of common envelopes. Future works will help clarifying the rich 
phenomenology related to this phenomena.

\section*{Acknowledgements}
We thank N. Fraija, P. Kumar, D. Lazzati, M. MacLeod, A. Murguia-Berthier, A. Raga, E. Ramirez-Ruiz and N. Soker for useful discussions. F.D.C. thanks the UNAM-PAPIIT grants IA103315 and IN117917. D.L.C. acknowledges support from C\'atedras CONACyT-Instituto de Astronom\'ia (UNAM).


\bsp	
\label{lastpage}
\end{document}